\documentclass[12pt]{article}
\pdfoutput=1
\usepackage{amsmath}
\usepackage[dvips]{graphicx}
\DeclareGraphicsExtensions{.png,.pdf}

\usepackage{color}
\input amssym

\def\red#1{{\color{red} #1}}

\begin{document}

\def\prg#1{\par\medskip\noindent{\bf #1}}  \def\ra{\rightarrow}
\newcounter{nbr}
\def\note#1{\bitem\vspace{-5pt}\addtocounter{nbr}{1}
            \item{} #1\vspace{-5pt}
            \eitem}
\def\lra{\leftrightarrow}              \def\Ra{\Rightarrow}
\def\nin{\noindent}                    \def\pd{\partial}
\def\dis{\displaystyle}                \def\Lra{{\Leftrightarrow}}
\def\grl{{GR$_\Lambda$}}               \def\vsm{\vspace{-8pt}}
\def\cs{{\scriptstyle\rm CS}}          \def\ads3{{\rm AdS$_3$}}
\def\Leff{\hbox{$\mit\L_{\hspace{.6pt}\rm eff}\,$}}
\def\bull{\raise.25ex\hbox{\vrule height.8ex width.8ex}}
\def\ric{{Ric}}                        \def\tmgl{\hbox{TMG$_\Lambda$}}
\def\Lie{{\cal L}\hspace{-.7em}\raise.25ex\hbox{--}\hspace{.2em}}
\def\sS{\hspace{2pt}S\hspace{-0.83em}\diagup}
\def\hd{{^\star}}                      \def\dis{\displaystyle}
\def\mb#1{\hbox{{\boldmath $#1$}}}     \def\kn#1{\hbox{KN$#1$}}
\def\ul#1{\underline{#1}}              \def\phb{\phantom{\Big|}}
\def\nb{~\marginpar{\bf\Large ?}}      \def\ph{\phantom{xxx}}

\def\hook{\hbox{\vrule height0pt width4pt depth0.3pt
\vrule height7pt width0.3pt depth0.3pt
\vrule height0pt width2pt depth0pt}\hspace{0.8pt}}
\def\inn{\hook}
\def\first{\rm (1ST)}  \def\second{\hspace{-1cm}\rm (2ND)}
\def\ppl{{pp${}_\Lambda$}}

\def\G{\Gamma}        \def\S{\Sigma}        \def\L{{\mit\Lambda}}
\def\D{\Delta}        \def\Th{\Theta}       \def\Ups{\Upsilon}
\def\a{\alpha}        \def\b{\beta}         \def\g{\gamma}
\def\d{\delta}        \def\m{\mu}           \def\n{\nu}
\def\th{\theta}       \def\k{\kappa}        \def\l{\lambda}
\def\vphi{\varphi}    \def\ve{\varepsilon}  \def\p{\pi}
\def\r{\rho}          \def\Om{\Omega}       \def\om{\omega}
\def\s{\sigma}        \def\t{\tau}          \def\eps{\epsilon}
\def\nab{\nabla}      \def\btz{{\rm BTZ}}   \def\heps{{\hat\eps}}

\def\bR{\bar{R}}      \def\bT{\bar{T}}     \def\hT{\hat{T}}
\def\tG{{\tilde G}}   \def\cF{{\cal F}}    \def\cA{{\cal A}}
\def\cL{{\cal L}}     \def\cM{{\cal M }}   \def\cE{{\cal E}}
\def\cH{{\cal H}}     \def\hcH{\hat{\cH}}  \def\cT{{\cal T}}
\def\hA{\hat{A}}      \def\hB{\hat{B}}     \def\hK{\hat{K}}
\def\cK{{\cal K}}     \def\hcK{\hat{\cK}}  \def\cT{{\cal T}}
\def\cO{{\cal O}}     \def\hcO{\hat{\cal O}} \def\cV{{\cal V}}
\def\tom{{\tilde\omega}}  \def\cE{{\cal E}} \def\bH{\bar{H}}
\def\cR{{\cal R}}    \def\hR{{\hat R}{}}   \def\hL{{\hat\L}}
\def\tb{{\tilde b}}  \def\tA{{\tilde A}}   \def\hom{{\hat\om}}
\def\tT{{\tilde T}}  \def\tR{{\tilde R}}   \def\tcL{{\tilde\cL}}
\def\he{{\hat e}}    \def\hom{{\hat\om}}   \def\hth{\hat\theta}
\def\hxi{\hat\xi}    \def\hg{\hat g}       \def\hb{{\hat b}}
\def\tH{{\tilde H}}  \def\tV{{\tilde V}}   \def\bA{\bar{A}}
\def\bV{\bar{V}}     \def\bxi{\bar{\xi}}
\def\knl{\text{KN}$(\l)$}   \def\mknl{\text{KN}\mb{(\l)}}
\def\bPhi{\bar\Phi}
\def\chm{\checkmark}                \def\chmr{\red{}}
\vfuzz=2pt 
\def\nn{\nonumber}
\def\be{\begin{equation}}             \def\ee{\end{equation}}
\def\ba#1{\begin{array}{#1}}          \def\ea{\end{array}}
\def\bea{\begin{eqnarray} }           \def\eea{\end{eqnarray} }
\def\beann{\begin{eqnarray*} }        \def\eeann{\end{eqnarray*} }
\def\beal{\begin{eqalign}}            \def\eeal{\end{eqalign}}
\def\lab#1{\label{eq:#1}}             \def\eq#1{(\ref{eq:#1})}
\def\bsubeq{\begin{subequations}}     \def\esubeq{\end{subequations}}
\def\bitem{\begin{itemize}}           \def\eitem{\end{itemize}}
\renewcommand{\theequation}{\thesection.\arabic{equation}}

\title{Memory effect of the pp waves with torsion}

\author{B. Cvetkovi\'c and D. Simi\'c\footnote{
        Email addresses: {\tt cbranislav@ipb.ac.rs,
                          dsimic@ipb.ac.rs}}\\
Institute of Physics, University of Belgrade \\
                      Pregrevica 118, 11080 Belgrade, Serbia}
\date{\today}
\maketitle

\begin{abstract}
We analyse the motion of test particles in the spacetime  of the plane-fronted (pp) waves with torsion in four-dimensions. We conclude that there is a velocity memory effect in the direction of advanced time  and along radial direction,  while we have  rotation of  particles in angular direction. The velocity memory effect in the aforementioned directions is severely affected by the value of the tordion mass and  probably it is not observable. A very interesting, probably observable effect, stemms from the rotation, which is insensitive to the tordion mass.

\end{abstract}

\section{Introduction}
\setcounter{equation}{0}

When a gravitational wave passes through a system  of test particles it induces an observable disturbance of the system \cite{1,2}. In other words a system remembers that a wave passed, and for this reason it is known as the memory effect.

There are two possible outcomes when a wave passes, disregarding the trivial possibility  that everything is translated or boosted in the same way which is  unobservable. The first scenario is that relative velocity of test particles is zero while they suffer a permanent displacement depending on their initial conditions. This is known as the displacement memory effect \cite{1,2}, for the results on nonlinear contribution to the memory effect see  \cite{3,4}. The appearance of  displacement memory effect is questioned in \cite{5,6}, where  authors concluded that test particles will have non-zero relative velocity. This variation of the memory effect is known as the velocity memory effect, for recent development see  \cite{7}.

All of the previously mentioned results are obtained in the framework of general relativity and memory effect is not much investigated beyond it. Some of the results, known to the authors, are memory effect for massive graviton investigated in Ref. \cite{8}, while, the memory effect of the gravitational waves with torsion in  the Poincar\'e gauge theory (PGT)  has been investigated only in three-dimensions in Ref. \cite{9}.

The aim of this paper is  to fill the gap in the literature, namely to extend results about the memory effect to the gravitational waves with torsion in four-dimensions.

Basic dynamical variables in PGT  \cite{10,11,12} are the
tetrad field $b^i$ and the Lorentz connection $\om^{ij}=-\om^{ji}$
(1-forms), and the associated field strengths are the torsion $T^i=d
b^i+\om^i{_k}\wedge b^k$ and the curvature
$R^{ij}=d\om^{ij}+\om^i{_k}\wedge\om^{kj}$ (2-forms). By construction, PGT
is characterized by a Riemann-Cartan geometry of spacetime, and its
physical content is directly related to the existence of mass and spin as
basic characteristics of matter at the microscopic level.
General PGT Lagrangian $L_G$ is at most quadratic in the field strengths.
The number of independent (parity invariant) terms in $L_G$ is nine, which
makes the corresponding dynamical structure rather complicated.

The paper is organized  as follows. First, we review the gravitational pp wave solutions with torsion in four-dimensions. After that, we derive the geodesic equations in this pp wave spacetime. We finally numerically  solve the geodesic equations.

Our conventions are as follows. The Latin indices $(i, j, ...)$ refer to
the local Lorentz (co)frame and run over $(0,1,2,3)$, $b^i$ is the tetrad
(1-form), $h_i$ is the dual basis (frame), such that $h_i\inn b^k=\d^i_k$;
the volume 4-form is $\heps=b^0\wedge b^1\wedge b^2\wedge b^3$, the Hodge
dual of a form $\a$ is $\hd\a$, with $\hd 1=\heps$, totally antisymmetric
tensor is defined by $\hd(b^i b^jb^kb^l)=\ve^{ijkl}$
and normalized to $\ve_{0123}=+1$; the exterior product of forms is
implicit in all expressions.

\section{Review of the pp waves}
\setcounter{equation}{0}

In this section, we give an overview of 4D pp waves in PGT. For details see \cite{13}.
\subsection{pp waves without torsion}

\subsubsection{Geometry}

In local coordinates $x^\m=(u,v,y,z)$, the metric of the
pp waves is of the form

\be
ds^2=du(Hdu+2dv)-(dy^2+dz^2)\, ,     \lab{2.1a}
\ee
where  the unknown metric function $H=H(u,y,z)$ is to be obtained from the field equations.
The advanced time $v$ is an affine parameter along the null geodesics $x^\m=
x^\m(v)$, and $u$ is retarded time such that $u=$ const. are the spacelike
surfaces parameterized by $x^\a=(y,z)$. Since the null vector
$\xi=\xi(u)\pd_v$ is orthogonal to these surfaces, they are regarded as
wave surfaces, and $\xi$ is the null direction (ray) of the wave
propagation.

We choose the tetrad field (coframe) to be of the form
\bsubeq\lab{2.2}
\be
b^0:=du\, ,\qquad b^1:=\frac H2du+dv\, ,\qquad
b^2:=dy\, ,\qquad b^3:=dz\, ,
\ee
so that $ds^2=\eta_{ij}b^i\otimes b^j$, where $\eta_{ij}$ is the half-null
Minkowski metric:
$$
\eta_{ij}=\left( \ba{cccc}
             0 & 1 & 0 & 0  \\
             1 & 0 & 0 & 0  \\
             0 & 0 & -1& 0  \\
             0 & 0 & 0 & -1
                \ea
       \right)\, .
$$
The corresponding dual frame $h_i$ is given by
\be
h_0=\pd_u-\frac H2\pd_v\, ,\qquad h_1=\pd_v\, ,
\qquad h_2=\pd_y\, ,\qquad h_3=\pd_z\, .
\ee
\esubeq
For the coordinates $x^\a=(y,z)$ on the wave surface, we have:
$$
x^c=b^c{_\a}x^\a=(y,z)\, ,\qquad
\pd_c=h_c{^\a}\pd_\a=(\pd_y,\pd_z)\, ,
$$
where $c=2,3$. After introducing the notation $i=(A,a)$, where $A=0,1$ and $a=(2,3)$, one can
find the compact form of the Riemannian connection $\om^{ij}$:
\bea \lab{2.3}
&&\om^{Ac}=\frac12k^Ab^0\pd^c H \, ,
\eea
where $k^i=(0,1,0,0)$ is a null propagation vector, $k^2=0$.

The above connection defines the Riemannian curvature $R^{ij}=d\om^{ij}+\om^i{_m}\om^{mj}$; for $i<j$, it is given by
\bsubeq\lab{2.4}
\be
R^{ij}=2b^0k^{[i}Q^{j]}
\ee
where $Q^c$ is a 1-form introduced by Obukhov \cite{14},
\be
Q^2=\frac12\pd_{yy}H b^2
     +\frac12\pd_{yz}Hb^3,
\qquad Q^3=\frac12\pd_{zz}Hb^3+\frac12\pd_{yz}H b^2\,.
\ee
\esubeq
The Ricci 1-form $\ric^i:=h_m\inn\ric^{mi}$ is given by
\bea
&&\ric^i=b^0k^i Q\, ,                                  \nn\\
&&Q=h_c\inn Q^c=\frac{1}{2}\left[\pd_{yy}H+\pd_{zz}H\right],         \lab{2.5}
\eea
and the scalar curvature $R:=h_i\inn\ric^i$ vanishes.

\subsubsection{pp waves in GR}

Starting with the action $I_0=-\int d^4 xa_0R$, one can derive the
GR field equations in vacuum:

\be
2a_0G^n{_i}=0\, ,
\ee
where $G^n{_i}$ is the Einstein tensor.
As a consequence, the metric function $H$ must obey
\be
\pd_{yy}H+\pd_{zz}H=0\, .                      \lab{2.8}
\ee
There is a simple solution of these equations,
\be
H_c=A(u)+B_\a(u)x^\a\, ,
\ee
for which $Q^a$ vanishes. This solution is trivial (or pure gauge), since
the associated curvature takes the background form, $R^{ij}=0$.

\subsection{pp waves with torsion}\label{sec3}

\subsubsection{Geometry of the ansatz}

We assume that the form of the triad field \eq{2.2} remains unchanged, while looking at the Riemannian connection \eq{2.3}, one can notice that its
radiation piece appears only in the $\om^{1c}$ components:
$$
(\om^{1c})^R=\frac{1}{2}(h^{c\a}\pd_\a H) b^0\, .
$$
This motivates us to construct a new connection by applying the rule
\bsubeq\lab{2.9}
\be
\frac12\pd_\a H\to \frac12\pd_\a H+K_\a\,, \qquad K_\a=K_\a(u,y,z)\, ,
\ee
where $K_\a$ is the component of the 1-form $K=K_\a dx^\a$ on the wave
surface. Thus, the new form of $(\om^{ij})^R$ reads
\be
(\om^{ic})^R:=k^ih^{c\a}(\frac12\pd_\a H+K_\a)\,b^0\, ,.
\ee
\esubeq

The geometric content of the new connection is found by calculating the
torsion:
\be
T^i=\nab b^i+\om^i{_m}b^m=k^ib^0(b^2 K_y+b^3K_z)=k^ib^0b^cK_c\, .                             \lab{3.2}
\ee
The only nonvanishing irreducible piece of $T^i$ is ${}^{(1)}T^i$.

The new connection modifies also the curvature, so that its radiation
piece becomes
\bsubeq
\be
(R^{1c})^R=k^1b^0\Om^c\, ,\qquad \Om^c:=Q^c+\Th^c\, ,
\ee
where the term $\Th^c$ that represents the contribution of torsion is
given by
\bea
\Th^2=\pd_y K_yb^2-\pd_z K_yb^3\,,\qquad \Th^3=\pd_z K_z b^3-\pd_y K_zb^2.\nn
\eea
The covariant form of the curvature reads
\be
R^{ij}=2b^0k^{[i}\Om^{j]}\, ,                     \lab{2.11b}
\ee
and the Ricci curvature takes the form
\be
\ric^i=b^0k^i\Om\, ,\qquad \Om:=h_c\inn\Om^c=Q+\Th\,.
\ee
The torsion has no influence on the scalar curvature and it again vanishes.
\esubeq
Thus, our ansatz defines a RC geometry of spacetime.

\subsubsection{Massive torsion waves}
\setcounter{equation}{0}

The irreducible decomposition of the curvature implies (see \cite{13})
\be
{}^{(3)}R_{ij}=0\, ,\qquad {}^{(5)}R_{ij}=0\, ,\qquad {}^{(6)}R_{ij}=0
\ee
whereas the remaining pieces ${}^{(n)}R^{ij}$ are defined by their
nonvanishing components as
\bsubeq\lab{3.5}
\bea
&&{}^{(2)}R^{1c}=\frac{1}{2}\hd(\Psi^1b^c)\, ,\qquad
  {}^{(4)}R^{1c}=\frac{1}{2}(\Phi^1b^c)\, ,                     \nn\\
&&{}^{(1)}R^{1c}
  =b^0\left(\Om^{(ce)}-\frac{1}{2}\eta^{ce}\Om\right)b_e\, ,    \lab{3.5a}
\eea
where the 1-forms $\Phi^i$ and $\Psi^i$ are given by
\bea
&&\Phi^i=k^i b^0(Q+\Th)\, ,\qquad
  \Th=\pd_y K_y
             +\pd_z K_z\, ,       \nn\\                                       \nn\\
&&\Psi^i=X^i=-k^i b^0\S\, ,\qquad
  \S=\pd_z K_y
            -\pd_y K_z \, .
\eea
\esubeq
Having found ${}^{(1)}T_i$ and ${}^{(n)}R_{ij}$, we obtain the following form of the two
PGT field equations \cite{13}:
\bsubeq\lab{3.6}
\bea
\text{(1ST)}
&&\L=0\,,\qquad a_1\Th-a_0(Q+\Th)=0\, ,                 \lab{3.6a}\\[3pt]
\text{(2ND)}
&&-(b_2+b_1)(\nab\Psi^1)b^2-(b_4+b_1)(\nab\Phi^1)b^3
                 -2\bigl(a_0-A_1\bigr)T^1b^3=0\, ,\quad         \nn\\
&&-(b_2+b_1)(\nab\Psi^1)b^3+(b_4+b_1)(\nab\Phi^1)b^2
                 +2\bigl(a_0-A_1\bigr)T^1b^2=0\,.\quad         \lab{3.6b}
\eea
\esubeq

Leaving (1ST) as it is, (2ND) can be given a more clear structure as follows:
\bsubeq\lab{3.8}
\bea
&&(\pd_{yy}+\pd_{zz})\Th
                   -m_{2^+}^2\Th=0\,,
       \quad\dis m_{2^+}^2:=\frac{2a_0(a_0-a_1)}{a_1(b_1+b_4)},  \lab{3.8a}\\[3pt]
&&(\pd_{yy}+\pd_{zz})\S
                   -m_{2^-}^2\S=0\,,
       \quad\dis m_{2^-}^2:=\frac{2(a_0-a_1)}{b_1+b_2}.          \lab{3.8b}
\eea
\esubeq
The parameters $m^2_{2^\pm}$ have a simple physical interpretation. They represent masses of the spin-$2^\pm$ torsion modes
with respect to the $M_4$ background \cite{15},
$$
\bar m^2_{2^+}=\frac{2a_0(a_0-a_1)}{a_1(b_1+b_4)}\, ,\qquad
\bar m_{2^-}^2=\frac{2(a_0-a_1)}{b_1+b_2}\, .
$$

In $M_4$, the physical torsion modes are required to satisfy the
conditions of no ghosts (positive energy) and no tachyons (positive $m^2$)
\cite{15,16}. However, for spin-$2^+$ and spin-$2^-$ modes, the
requirements for the absence of ghosts, given by the conditions
$b_1+b_2<0$ and $b_1+b_4>0$, do not allow for both $m^2$ to be positive.
Hence, only one of the two modes can exist as a propagating mode (with
finite mass), whereas the other one must be ``frozen" (infinite mass).

Important point to be noted is that the two spin-$2$ sectors have very different
dynamical structures.
\bitem
\item[$-$] In the spin-$2^-$ sector, the infinite mass of the spin-$2^+$
    mode implies $\Th=0$, while (1ST) gives $Q=0$, which is nothing other than
    the GR field equation for metric. Consequently, the presence of torsion
    has no influence on the metric.\vsm
\item[$-$] In the spin-$2^+$ sector, the infinite mass of the spin-$2^-$
    mode leads to $\S=0$, whereas (1ST) gives that $Q$ is proportional to
    $\Th$, with $\Th\ne 0$. Leading to the conclusion that the torsion function $\Th$ has a
    decisive dynamical influence on the metric.
\eitem

We shall focus our attention on the spin-$2^+$ sector, where the metric appears to be a genuine dynamical effect of PGT.

\subsubsection{Solutions in the spin-$2^+$ sector}

After  introducing polar coordinates
$y=\r\cos\vphi,z=\r\sin\vphi$, equation \eq{3.8a} takes the form
\bsubeq\lab{4.2}
\be
\left(\frac{\pd^2}{\pd\r^2}+\frac{1}{\r}\frac{\pd}{\pd \r}
+\frac{1}{\r^2}\frac{\pd^2}{\pd\vphi^2}\right)\Th-m^2\Th=0\, .
\ee
Looking for a solution of $\Th$ in the form of a Fourier expansion,
$$
\Th=\sum_{n=0}^\infty \Th_n(\r)(c_ne^{in\vphi}+\bar c_n e^{-in\vphi})\, ,
$$
we obtain:
\be
\Th_n''+\frac{1}{\r}\Th_n'
     -\left(\frac{n^2}{\r^2}+m^2\right)\Th_n=0\, ,    \lab{4.2b}
\ee
\esubeq
where prime denotes $d/d\r$.

The general solution of Eq. \eq{4.2b} has the form
\be
\Th_n=c_{1n}J_n(-im\r)+c_{2n}Y_n(-im\r)\, ,\qquad n=0,1,2,\dots  \lab{4.5}
\ee
where $J_n$ and $Y_n$ are Bessel functions of the 1st and 2nd kind,
respectively.

\subsubsection{Solutions for the metric function \mb{H}}

For a given $\Th$, the first PGT field equation
$a_0Q=(a_1-a_0)\Th$, with $Q$ defined in \eq{2.5}, represents a
differential equation for the metric function $H$:
\be
(\pd_{yy}+\pd_{zz})H=\frac{2(a_1-a_0)}{a_0}\Th\, .\lab{4.6}
\ee
This is a second order, linear nonhomogeneous differential equation, and
its general solution can be written as
$$
H=H^h+H^p\, ,
$$
where $H^h$ is the general solution of the homogeneous equation, and $H^p$
a particular solution of \eq{4.6}. One finds that there is a simple particular solution for $H$:
\bsubeq\lab{4.7}
\be
H^p=\s V\, ,\qquad \s=\frac{2(a_1-a_0)}{m^2a_0}\, .
\ee
On the other hand, $H^h$ coincides with the general vacuum solution of
\grl, see \eq{2.8}. Since our idea is to focus on the genuine torsion
effect on the metric, we choose $H^h=0$ and adopt $H^p$ as the most
interesting PGT solution for the metric function $H$. Thus, we have
\be
H_n=\s \Th_n\, .
\ee
\esubeq
The solutions  for torsion functions are given in Appendix A.


\section{Geodesic motion}
\setcounter{equation}{0}

In this section we shall examine the {\it geodesic motion} of particles in the field of the massive gravitational wave with torsion.
We shall consider the motion of spinless particles in a gravitational field, which follow geodesic lines. It is known that torsion  affects the motion of the particles with spin by causing its precession \cite{12,17}. However,  the gravitational waves with torsion, which we are considering  are intrinsically  different from  the well-known spherically symmetric (static or stationary) solutions
of  PGT \cite{18} (for review see \cite{19}). The  metric of these spherically symmetric  solutions is "independent" of torsion in the sense that it represents Schwarzschild (or Schwarzschild AdS, Kerr etc.)  metric and the motion of spinless particles is not affected by the presence torsion. For the gravitational wave solution \eq{4.2}, metric crucially depends on torsion, as we noted in the previous section.

\prg{Christoffel  connection.} The non-vanishing components of Christoffel  connection  in polar coordinates are given by
\bea
&&\tilde{\G}^v{}_{uu}=\frac12\pd_u H\,,\qquad\tilde{\G}^v{}_{u\r}=\frac12H'\,,\qquad \tilde\G^v{}_{u\varphi}=\frac12\pd_\vphi H\,,\nn\\
&&\tilde\G^\r{}_{uu}=\frac12H'\,,\qquad \tilde\G^\r_{\vphi\varphi}=-\r\,,\nn\\
&&\tilde\G^\vphi{}_{uu}=\frac{1}{2\r^2}\pd_\varphi H\,,\qquad \tilde\G^\vphi{}_{\r\vphi}=\frac{1}{\r}\,,
\eea
where $H':=\pd_\r H$.

Let us mention that we shall consider the solution with non-trivial contribution to metric function (and consequently Christoffel connection) stemming from the presence of torsion.
\prg{Geodesic equations.} The geodesic equation for $u$ takes the expected form
\bea
\frac{d^2u}{d\l^2}=\ddot u=0\,.
\eea
Therefore without the loss of generality we can assume $u\equiv \l$.

The equation for $v$, $\rho$ and $\vphi$ are given by:
\bea
\ddot{v}+\frac12\pd_u H+H'\dot{\r}+\pd_\varphi H \dot{\varphi}=0\,.\lab{4.5}\\
\ddot{\r}+\frac 12 H'-\r \dot{\varphi}^2=0\,,\lab{4.3}\\
\ddot{\varphi}+\frac{1}{2\r^2}\pd_\varphi H+\frac{2}{\r}\dot{\r}\dot{\varphi}=0\,.\lab{4.4}
\eea

 We shall solve the geodesic equations numerically, but let us
 first make some reasonable simplifications.

Firsty,  $v$ appears only as a second derivative because $H$ is independent of it. Consequently, we have a shift symmetry
\be
v\rightarrow v+c_0+c_1u\,,
\ee
which means that initial conditions at time $u_i$ can be chosen as
\be
v[u_i]=v'[u_i]=0\,.
\ee
Secondly,  in the metric function $H$  there is  a factor
\be
\sigma=\frac{2(a_1-a_0)}{m^2a_0}\,,
\ee
where $a_0=\frac{1}{16\pi G}$ is coupling constant of general relativity and $a_1$ corresponds to  correction in the action stemming from torsion.
Experimental results suggest that $a_1$ is much smaller than $a_0$ so we can approximate
\be
\sigma \approx -\frac{2}{m^2}\,.
\ee
Also, we can introduce reduced variables
\bea
\tilde{v}=m^2v\,, & r=m\r\,,
\eea
while $\varphi$ remains the same. In these variables geodesic equations
do not have explicit dependence on $m$ and have more suitable form for numerical calculations.

\subsection{Memory effect}
We have one more unknown in geodesic equations and that is the form of functions $c_{1n}$ and $c_{2n}$. We expect that their exact form is not specially  important, as long as they sufficiently fast tend to zero at infinity. But, we encountered numerical problems because for polynomial fall-off the software  cannot handle the computational  complexity. Because, of this problem we decided to focus to Gaussian form of functions, more precisely to the form $e^{-(u-5)^2}$. For the initial time we chose $u=0$. As we already noted the initial conditions for $\tilde{v}$ are
\be
v[0]=\dot{v}[0]=0\,,
\ee
and we assume that particle is initially at rest
\be
\dot{\r}[0]=\dot{\varphi}[0]=0\,.
\ee
So, the only variable inputs are $\r[0]$ and $\varphi[0]$ as well as   the modes $c_{1n}$ and $c_{2n}$ we are including.
\prg{Mode $J_0$.}
In this case we set $H=J_0(-ir)e^{-(u-5)^2}$. Because nothing explicitly depends of $\varphi$ it remains the same as at initial time. In Fig. 1. we plot radial velocity $\dot{r}$ in function of $u$. While the Fig. 2. shows the value of velocity $\dot{v}$.
\begin{center}
\begin{tabular}{c}
\includegraphics[height=4cm]{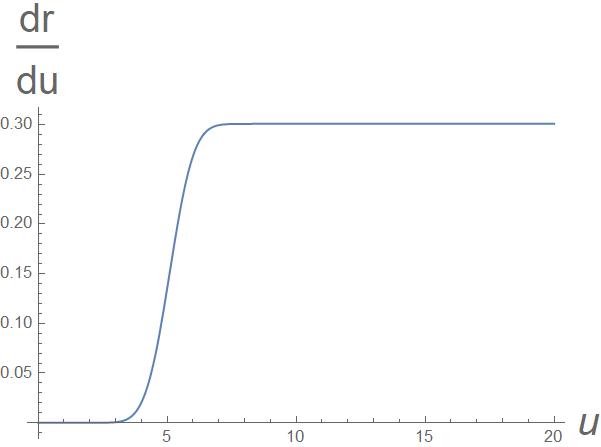}\hspace{1cm}\includegraphics[height=4cm]{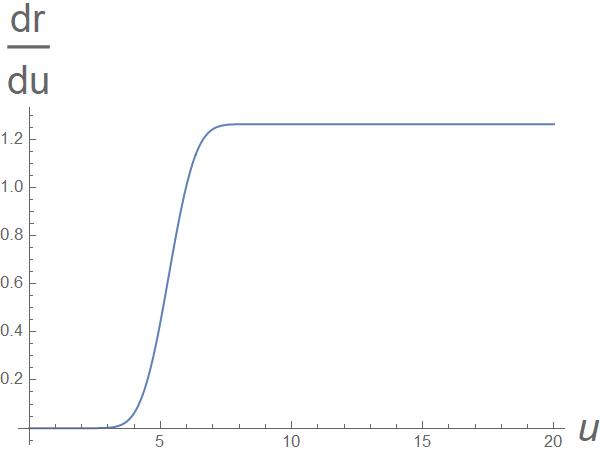}\\
Figure 1: The plot for the particle velocity $\dot r$ for $r[0]=1$ and $r[0]=2$\,.
\end{tabular}
\begin{tabular}{c}
	\includegraphics[height=4cm]{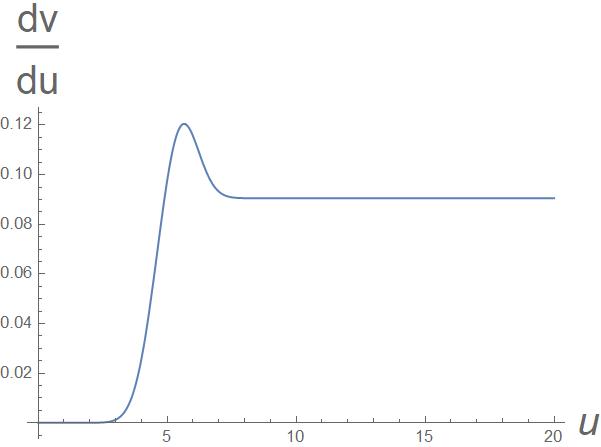}\hspace{1cm}\includegraphics[height=4cm]{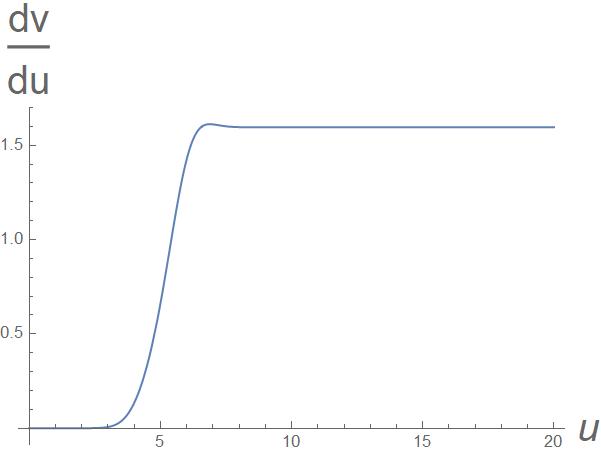}\\
Figure 2: The plot for the particle velocity $\dot {\text{v}}$ for $r[0]=1$ and $r[0]=2$\,.
\end{tabular}
\end{center}

\prg{Mode $J_2$.}
In this case we set $H=J_2(-ir)e^{-(u-5)^2}\sin(2\varphi)$. In Fig. 3. we plot the radial velocity $\dot{r}$. In Fig. 4. we show the value of angle $\varphi$. We see that in angular direction we have displacement memory effect in contrary to the others where we have velocity memory effect.
\begin{center}
\begin{tabular}{c}
	\includegraphics[height=4cm]{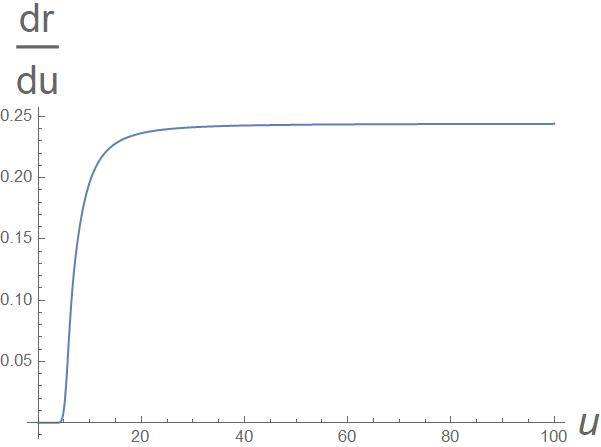}\hspace{1cm}\includegraphics[height=4cm]{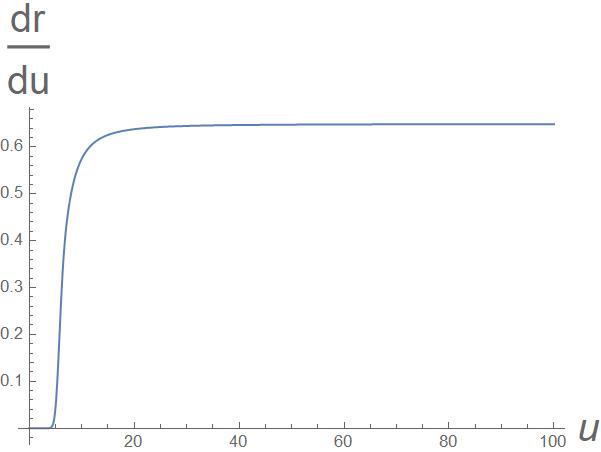}\\
Figure 3: The plot for the particle velocity $\dot r$ for $r[0]=1$ and $r[0]=2$ in both plots $\varphi[0]=0$\,.
\end{tabular}

\begin{tabular}{c}
	\includegraphics[height=4cm]{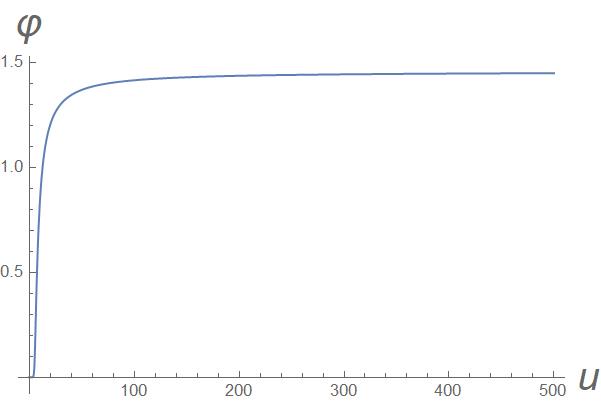}\hspace{1cm}\includegraphics[height=4cm]{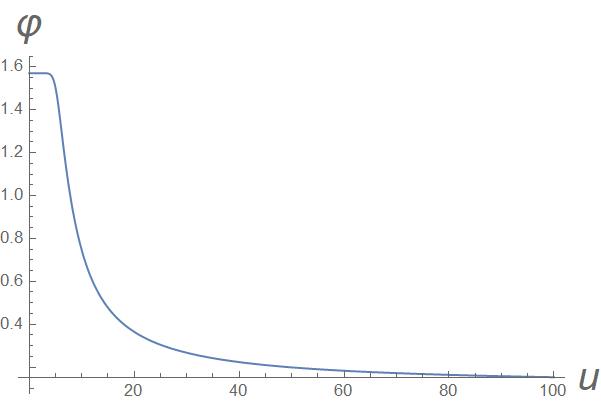}\\
Figure 4: The plot for the particles angular position $\varphi$ for $\varphi[0]=0$ and $\varphi[0]=\frac{\pi}{2}$ in both plots $r[0]=1$\,.
\end{tabular}
\end{center}
\prg{Mode $J_4$.} In this case we set $H=J_4(-ir)e^{-(u-5)^2}\sin(4\varphi)$.  In Fig. 5. we plot the radial velocity $\dot{r}$. In Fig. 6. we show the value of angle $\varphi$.
\begin{center}
\begin{tabular}{c}
	\includegraphics[height=4cm]{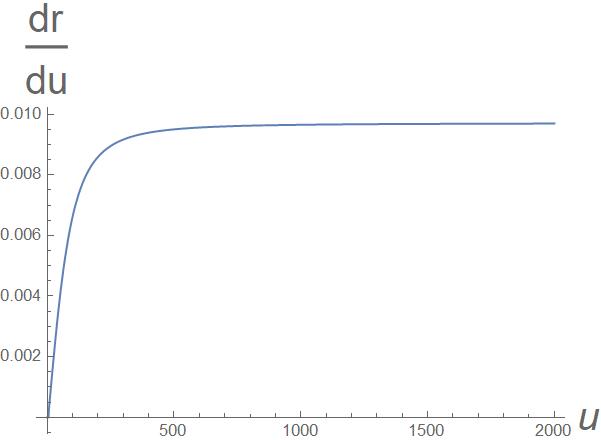}\hspace{1cm}\includegraphics[height=4cm]{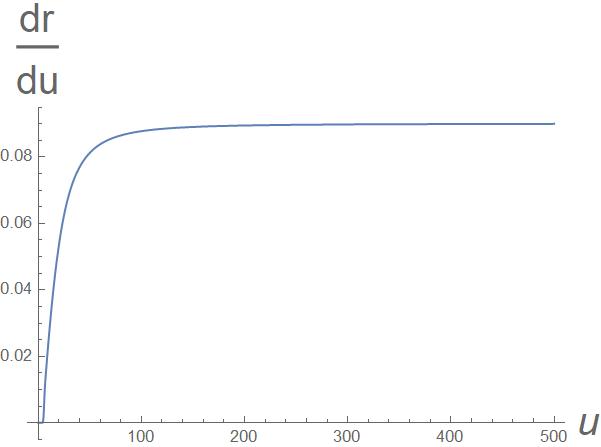}\\
Figure 5: The plot for the particle velocity $\dot r$ for $r[0]=1$ and $r[0]=2$ in both plots $\varphi[0]=0$\,.
\end{tabular}

\begin{tabular}{c}
\includegraphics[height=4cm]{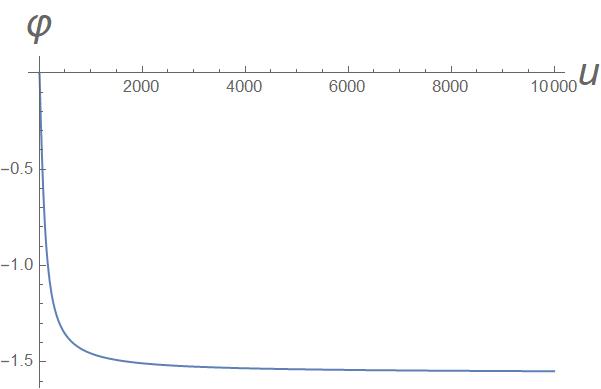}\hspace{1cm}\includegraphics[height=4cm]{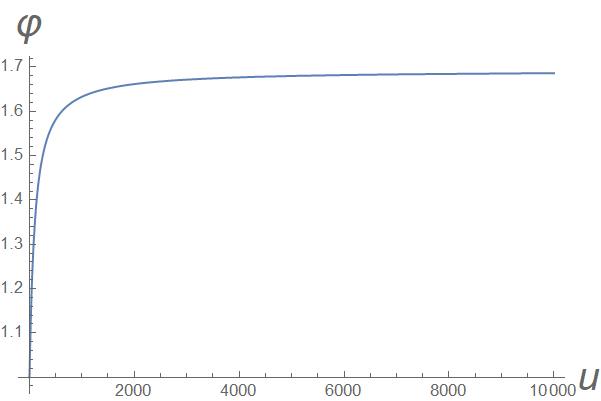}\\
Figure 6: The plot for the particles angular position  $\varphi$ for $\varphi[0]=0$ and $\varphi[0]=1$ in both plots $r[0]=1$\,.
\end{tabular}
\end{center}

\section{Discussion}
\setcounter{equation}{0}
We studied the geodesic motion of test particle in the presence of the pp wave with torsion. Our analysis discovered that particles, after the passage of the wave, show a combination of displacement and velocity memory effect. In the
angular direction we discovered that pp waves induce displacement memory effect. Let un note, for comparison, that velocity memory
effect takes place for axial gravitational waves \cite{20}. After the passage of axial wave burst particles rotate with constant
angular velocity around the symmetry axis.

Because we defined new variables $\tilde{v}=m^2v$ and $r=m\r$ our results have to be interpreted with taking into account the reasonable value of $m$. According to the  CERN results we expect that possible mass of tordion is no less that $10\, \rm TeV$
\be
m\geq 10\, \rm TeV\,.
\ee
This order of magnitude of the mass is equivalent to the length scale $\ell_m$
\be
\ell_m \approx 10^{-20}\,,\rm m\,.
\ee
Due to the very large mass of tordion or, equivalently, very small length scale the physical values of $v$ and $\r$ are very small and probably not observable. Fortunately $\varphi$ is insensitive to the value of the mass and offers a possible observable effect. We see from Figs. 4 and  6 that depending on the initial angular position the particle will be rotated by a different angle. Consequently, the particles initially set at some positions on a circle will be rotated relatively to each other. This is the possible experimental setup for the detection of torsion waves.

\section*{Acknowledgments}

This work was partially supported by the Serbian Ministry of Education, Science and Technological development.

\appendix

\section{Solutions for the torsion functions \mb{K_\a}}
\setcounter{equation}{0}

In the spin-$2^+$ sector, the torsion functions $K_\a$ can be determined
(by using the condition $\S=0$) from the equations:
\be
\pd_y \Th+m^2K_y=0\,,\qquad
\pd_z \Th+m^2K_z=0\, .                                 \lab{A.1}
\ee
Going over to polar coordinates,
$$
K_y=K_\r\cos\vphi-\frac{K_\vphi}{\r}\sin\vphi\, ,\qquad
K_z=K_\r\sin\vphi+\frac{K_\vphi}{\r}\cos\vphi\, ,
$$
the previous equations are transformed
into
\bsubeq
\be
K_\r=-\frac{1}{m^2}\pd_\r V\, ,\qquad
K_\vphi=-\frac{1}{m^2}\pd_\vphi V\, ,
\ee
or equivalently, in terms of the Fourier modes,
\be
K_{\r n}=-\frac{1}{m^2}\frac{p}{q}\pd_\r \Th_n\, ,\qquad K_{\vphi
n}=-\frac{1}{m^2}n\Th_n\, ,
\ee
\esubeq
where $K_\vphi=\sum_{n=1}^\infty (d_n e^{in\vphi}+\bar d_n e^{-i n\vphi})$
with $d_n=-ic_n$, and similarly for $K_\r$.

\end{document}